\documentclass{jpp}
\usepackage{graphicx}
\usepackage{epstopdf}

\usepackage[utf8]{inputenc}
\usepackage[T1]{fontenc}
\usepackage{amsmath}
\usepackage{subfig}
\usepackage{natbib}
\usepackage{mathtools,tikz}
\DeclareRobustCommand\sampleline[1]{%
  \tikz\draw[#1] (0,0) (0,\the\dimexpr\fontdimen22\textfont2\relax)
  -- (2em,\the\dimexpr\fontdimen22\textfont2\relax);%
}
\usepackage{hyperref} 
	\hypersetup{colorlinks=true,
		pdfstartview=FitV,
		linkcolor=blue,
		citecolor=blue,
		urlcolor=blue,
	}

\shorttitle{Kinetic Modelling of Runaway Electrons}
\shortauthor{K. Insulander Bj\"{o}rk et al.}

\title{Kinetic modelling of runaway electron generation in argon-induced disruptions in ASDEX Upgrade}

\author{K.~Insulander Bj\"{o}rk\aff{1} \corresp{\email{klaraib@chalmers.se, ppg@ipp.mpg.de}},
  G.~Papp\aff{2},
  O.~Embreus\aff{1},
  L.~Hesslow\aff{1},
  T.~F\"{u}l\"{o}p\aff{1},
  O.~Vallhagen\aff{1},
  A.~Lier\aff{2},
  G.~Pautasso\aff{2},
  A.~Bock\aff{2},
  the ASDEX Upgrade Team\footnote{See the author list of ``H. Meyer {\em et al}, 2019 Nucl. Fusion \href{https://iopscience.iop.org/article/10.1088/1741-4326/ab18b8}{{\bf 59} 112014}''} \and
  the EUROfusion MST1 Team\footnote{See the author list of ``B. Labit {\em et al}, 2019 Nucl. Fusion \href{https://iopscience.iop.org/article/10.1088/1741-4326/ab2211}{{\bf 59} 086020}''}}

\affiliation{\aff{1} Chalmers University of Technology, Gothenburg, 412 96, Sweden
\aff{2}Max Planck Institute for Plasma Physics, Garching, Germany}

\begin{document}

\maketitle

\begin{abstract}

Massive material injection has been proposed as a way to mitigate the formation of a beam of relativistic runaway electrons that may result from a disruption in tokamak plasmas. In this paper we analyse runaway generation observed in eleven ASDEX Upgrade discharges where disruption was triggered using massive gas injection. We present numerical simulations in scenarios characteristic of on-axis plasma conditions, constrained by experimental observations, using a description of the runaway dynamics with self-consistent electric field and temperature evolution in two-dimensional momentum space and zero-dimensional real space.
We describe the evolution of the electron distribution function during the disruption, and show that the runaway seed generation is dominated by hot-tail in all of the simulated discharges. We reproduce the observed dependence of the current dissipation rate on the amount of injected argon during the runaway plateau phase. Our simulations also indicate that above a threshold amount of injected argon, the current density after the current quench depends strongly on the argon densities. This trend is not observed in the experiments, which suggests that effects not captured by 0D kinetic modeling -- such as runaway seed transport -- are also important.
\end{abstract}

\section{Introduction}
\label{sec:intro}

Disruptions in tokamak plasmas may lead to the formation of a beam of relativistic, so-called {\em runaway electrons} (RE), which has the potential to severely damage plasma-facing components \citep{Hender_2007}. For this reason, much effort is directed towards the development of schemes to avoid, limit or mitigate the formation of such a beam. One proposed measure is using Massive Material Injection (MMI) in the form of gas (Massive Gas Injection - MGI) or frozen pellets (Shattered Pellet Injection - SPI) to avoid or dissipate the runaway electrons \citep{HollmannDMS,lehnen15disruptions}, and its efficiency has been demonstrated in medium-sized tokamaks \citep{hollmann15measurement,Pautasso_2016,pazsoldan17spatiotemporal,carnevale18runaway,reux15runaway,esposito17runaway,coda2019physics,mlynar19runaway,pautasso20generation}. However, the plasma currents, temperatures and densities of future devices such as ITER will be significantly larger than what can be achieved in current experiments, and simulations are necessary to foresee the effectiveness of massive material injections for disruption mitigation under such conditions. 

As MGI is widely used in current devices, cases where runaways are formed in MGI-induced disruptions provide a valuable dataset to properly understand the physics of runaway electron formation and dissipation in such scenarios. To gain this understanding, theoretical models have been formulated and implemented in computational tools which can be used to model disruptions \citep{BreizmanAleynikov2017Review}. To be applicable for predictions for ITER and beyond, theoretical tools must first be validated against existing experimental data to ensure that they capture the relevant physics.

One such kinetic modelling tool is {\sc code} (COllisional Distribution of Electrons), briefly described in section \ref{sec:models} and in detail in the paper by \citet{Stahl-NF-2016}. This tool includes modelling of many phenomena important for the studied scenarios, in particular partial screening of partially ionized impurities, and was therefore chosen for the present investigations. The present paper discusses the comparison of {\sc code} simulations with experimental data obtained from the ASDEX Upgrade tokamak (AUG, section \ref{sec:AUG}). Runaway-generating disruptions are deliberately triggered in AUG to obtain relevant data for studies of the connected phenomena, and parameters important for modelling are measured by plasma diagnostics, as described in sections \ref{sec:Ar} and \ref{sec:TQ}.

In a disruption, the induced electric field accelerates electrons above a certain critical velocity to relativistic energies. The Dreicer runaway generation is the result of velocity space diffusion of the electrons into the runaway region due to small angle collisions \citep{dreicer1959}. Existing runaways can create new fast electrons through close collisions with bulk electrons (secondary generation). The latter leads to an exponential growth of the number of REs - an avalanche \citep{sokolov1979multiplication,RosenbluthPutvinski1997}.

In the case of sudden cooling (a thermal quench, TQ), fast electrons do not have time to thermalize, and a so-called hot-tail forms in the electron distribution. Hot-tail generation is the dominant primary generation when the TQ duration is shorter than the collision time at the runaway threshold velocity \citep{helanderhottail}. In future fusion devices, due to the higher initial plasma temperature, the collision time is not much shorter than the expected duration of the TQ, and therefore a sizable hot-tail RE seed is likely to be produced \citep{SmithHottail2005}. A simplified analytical model for hot-tail generation has been formulated by \cite{Smith-PoP-08} but comparisons with kinetic simulations show that this model underestimates the runaway density by an order of magnitude \citep{Stahl-NF-2016}, so to gain a quantitative understanding of hot-tail generation, kinetic simulations are needed.

The future scenario we desire to understand is a spontaneous disruption, mitigated by MMI. However, since spontaneous disruptions do not occur reproducibly, we instead model disruptions which were deliberately triggered by an MGI, resulting in a scenario which is similar to the desired scenario in the important aspect that a runaway current is formed and dissipated in the presence of partly ionized high-Z materials.

Such scenarios have been considered also in the recent paper by \cite{Linder-2020}, where the {\sc astra-strahl} 1.5D transport code \citep{Fable_2013,Dux_1999} was used, including reduced kinetic models for Dreicer and avalanche runaway generation \citep{Hesslow-NF-19, Hesslow-JPP-19}.  The modelling results presented here intend to assess the kinetic part of the process, namely to which extent {\sc code} captures the central aspects of runaway current formation and dissipation and what role hot-tail generation plays. For a full understanding of the disruption scenario, kinetic simulation must be combined with other tools, most importantly modelling spatial dynamics, 3D MHD evolution and the atomic physics needed to determine ionization states.

\section{Kinetic Modelling}
\label{sec:models}
The relativistic finite-difference Fokker-Planck solver {\sc code} \citep{Stahl-NF-2016} simulates electron dynamics in plasmas in 2D momentum space. The plasma is assumed to be homogenous, i.e.~it is a 0D simulation in real space and radial transport or instabilities are not modelled. We focus on modelling the evolution of the electron momentum distribution function on-axis, and will use experimental data representative of the conditions on the plasma axis to the extent possible. 

{\sc code} calculates the time-evolving electron distribution function under the influence of collisions, synchrotron radiation reactions and electric field acceleration. In the simulations presented here, collisions are modelled by a relativistic test particle operator (\citet{BraamsKarney1989}, detailed in \citet{Hesslow-JPP-19}) and a simplified large-angle collision operator \citep{RosenbluthPutvinski1997}. Screening of partially ionized impurities is taken into account according to the model described by \cite{HesslowJPP}.  Bremsstrahlung radiation losses were found to be negligible in the scenarios considered here, as well as the differences between the fully conservative and the simplified avalanche operator \citep{olaknockon2018}. 

The electric field $E$ and the plasma current density $j$ is calculated self-consistently throughout the simulation, according to  \citet{Ecrit}:
\begin{equation}
	E  =-\frac{L a^2}{2 R}\frac{{\rm d} j}{{\rm d} t},
    	\label{eq:selfcons}
\end{equation} 
where the inductance $L$ is given by
\begin{equation}
	L \approx \mu_0 R \left[ \ln \left(\frac{8 R}{a} \right)-2 +\frac{l_i}{2}\right].
    	\label{eq:inductance}
\end{equation} 
The major and minor radii, $R$ and $a$ respectively, are given in table \ref{tab:AUG} and $l_i$ is the internal inductance. The value of $l_i$ differs between discharges, but has a negligible effect on the simulation results. The difference in calculated final runaway current between the cases using $l_i$ = 0 and $l_i$ = 1.5 (which is a common estimate for $l_i$ in AUG) was less than 1\% for AUG discharge \#35408. This is expected, partly since the inductance differs by only 3\% between the two cases, and partly since, as will be shown later, the primary RE generation is dominated by the hot-tail mechanism which is not sensitive to the induced electric field.  A more accurate expression for the inductance is $L = \lvert \Psi_p/I_p \rvert$ \citep{boozer_2018}, where $\Psi_p$ is the poloidal magnetic flux and $I_p$ is the plasma current. To calculate the on-axis magnetic flux $\Psi_p$ requires the plasma current density profile, which we do not know with an accuracy that motivates the use of the more accurate expression.

Using the test-particle collision operator is computationally efficient but leads to the underestimation of the ohmic current by about a factor of two \citep{helander} and in a self-consistent calculation, this needs to be compensated for. As the conductivity obtained by the test-particle operator $\sigma_{\rm{CODE},tp}$ is proportional to the fully relativistic electric conductivity $\sigma_{\rm{BK}}$ \citep{BraamsKarney1989} for a wide range of effective charges and temperatures \citep{Hoppe-ASDEX-2020}, we can model the plasma current density as
\begin{equation}
	j = j_{\rm{CODE},tp} + \Delta \sigma E, 
\end{equation}
where $j_{\rm{CODE},tp}$ is the current density calculated by {\sc  code} using the test particle operator, and $\Delta
\sigma(Z_\mathrm{eff},T_e)=\sigma_{\rm{BK}}-\sigma_{\rm{CODE},tp}$. The calculation of the effective ionic charge $Z_\mathrm{eff}$ is described in section \ref{sec:zeff} and the free electron temperature $T_e$ in section \ref{sec:TQ}.

\section{Experimental data}
\label{sec:experimental}

\subsection{ASDEX Upgrade}
\label{sec:AUG}
In this paper we model ASDEX Upgrade discharges specifically designed for the study of runaway electron dynamics \citep{Pautasso_2016}. ASDEX Upgrade is a medium sized tokamak at the Max Planck Institute for Plasma Physics in Garching, Germany. The typical runaway electron scenario uses a low density (initial free electron density $n_{e0}\approx 3{\cdot}10^{19}~\mathrm{m}^{-3}$), inner wall limited, circular (elongation $\kappa \approx 1.1$), ECRH (Electron Cyclotron Resonance Heating) heated plasma terminated with an argon MGI. The argon is held in a 0.1 l chamber, at room temperature, before the injection valve is triggered at 1.000 s into the discharge. In the considered discharges, the pressure in the valve ranged from 0.11 bar to 0.9 bar. During the simulated time span, the plasma position remained stable radially as well as vertically.

Eleven discharges with similar plasma parameters but with different amounts of argon injected were selected for modelling. As the discharges -- which represent a scan of injected argon quantity -- were selected from a database spanning multiple years, the electron temperature $T_{e0}$ before the disruption varies in the dataset ($T_{e0} = 8.0 \pm 2.9$~keV), because of varying experimental conditions and occasional ECRH gyrotron trips over the years. In ten of the modelled discharges, a runaway current was formed, and discharge \#35400, in which no runaway current was formed, was added for comparison. Parameters that are common to all modelled discharges are shown in table \ref{tab:AUG} and an overview of the basic parameters for the eleven selected discharges is presented in table \ref{tab:discharges}.

\begin{table}
\centering
\caption{Common parameters of all modelled discharges. The initial current is slightly lower, 0.71 MA, in discharge \#31318.}
\label{tab:AUG}
\begin{tabular}{l r l l} 
Magnetic field 				& $B =$		& 2.5 								& T \\
Major radius					& $R =$		& 1.65								& m \\
Minor radius					& $a =$ 	& 0.50								& m \\
Initial plasma current			& $I_{p0} =$	& 0.76							& MA\\
\end{tabular}
\end{table}

The initial on-axis current density is used as input to {\sc code} when creating an initial electron momentum distribution function. The on-axis current density $j$ can be estimated using current density profiles obtained from equilibrium reconstructions by CLISTE \citep{cliste}. For the modelled discharges, the initial on-axis current density $j_0$ was estimated to approximately 1.2 MA/m$^2$. The initial plasma current $I_{p0}$ is very similar between all discharges, and equals 0.76 MA as listed in table \ref{tab:AUG}. The conversion factor between the estimated $j_0$ and the measured $I_{p0}$ is thus 0.76/1.2 = 0.63 m$^2$. Since this conversion factor has the unit of m$^2$, it will be referred to as an "effective area", $A_\mathrm{eff}$. Application of this conversion factor results in an initial on-axis current density of 1.21 MA/m$^2$ $\pm$ <1\% for all discharges except the very early discharge \#31318 in which the initial on-axis current density is 1.13 MA/m$^2$. This estimated conversion factor was applied in the simulations of all discharges and throughout the simulated time period. Since the pre-disruption plasma is inner wall limited, and the post-disruption runaway beam is surrounded by a low temperature companion plasma, the radial extent of the two stages is comparable. 

\begin{table}
\centering
\caption{Basic on-axis parameters of the eleven simulated discharges and the notation used in this paper. The injection pressure is expressed in bars in the 0.1 l injection volume. The injected number of Ar atoms is estimated assuming a gas temperature of 300 K. The initial free electron density is the value given by CO$_2$ interferometry (average over the first 1.5 ms after the argon valve trigger, i.e. before Ar penetration into the plasma) and the initial electron temperature by the ECE measurements.}
\label{tab:discharges}
\begin{tabular}{c c c c c} 
Discharge 	& Injection	& Injected number	& Initial free electron	& Initial electron   \\
number 		& pressure [bar]	& of Ar atoms [$10^{19}$]	& density [$10^{19}$m$^{-3}$]	& temperature [keV]  \\[3pt]
\# 		& $p_{\rm{Ar}}$		& $N_{\rm{Ar}}$		& $n_{e0}$			& $T_{e0}$  \\[3pt]
35400 & 0.11 & 26 & 3.4 & 5.6\\
35401 & 0.15 & 36 & 2.6 & 6.1\\
34149 & 0.20 & 48 & 3.0 & 5.7\\
34183 & 0.31 & 74 & 2.8 & 5.5\\
34140 & 0.31 & 74 & 2.6 & 5.8\\
34084 & 0.33 & 79 & 3.0 & 4.3\\
35649 & 0.39 & 94 & 2.6 & 6.2\\
35650 & 0.40 & 96 & 2.4 & 5.3\\
35408 & 0.50 & 120 & 2.4 & 6.0\\
33108 & 0.73 & 175 & 3.1 & 7.2\\
31318 & 0.90 & 216 & 2.1 & 10.8\\
\end{tabular}
\end{table}

\subsection{Densities} \label{sec:Ar}
The disruptions were triggered by injection of argon into the plasma \citep{Pautasso_2016, Fable-NF-16}. When penetrating into the plasma, the injected argon is partly ionized. The density of free electrons, as well as the density and charge states of the argon atoms, directly affect the collision operator, and hence the evolution of the electron momentum distribution function. Thus, these parameters must be estimated and used as input to the simulations.

\subsubsection{Free electron density}
The free electron density is measured by CO$_2$ interferometry. This method yields the line integrated free electron density along the line of sight of the interferometer. The measured value can be divided by an estimate of the chord length, i.e. the portion of the line of sight that passes through the plasma, and the resulting density value is an average over the estimated chord length. Thus, the density value obtained by this method is not fully representative of the on-axis conditions we intend to simulate.

Previous work by \citet{Fable-NF-16} indicates that the free electron density increases rapidly on the plasma edge, but remains constant on-axis until the MHD mixing event that also causes the plasma current $I_p$ to increase for about a millisecond just before starting to decay due to the increased plasma resistivity. The time of the onset and end of this current spike are referred to as $t_{\rm onset}$ and $t_{\rm end}$ respectively, where $t_{\rm end}$ is defined by the peak of the current spike, and $t_{\rm onset}$ is the time when the measured $I_p$ starts to increase, just before the current spike. We assume that the on-axis free electron density remains constant at the pre-injection value until $t_{\rm onset}$, and that after $t_{\rm end}$ the plasma is homogenous enough for the measured line-averaged free electron density to be representative of the on-axis value. The data is smoothed using the \texttt{rloess} algorithm in \textsc{matlab} \citep{matlab} to avoid numerical difficulties caused by the signal noise. Between $t_{\rm onset}$ and $t_{\rm end}$, the free electron density is assumed to increase linearly. Simulations were run with different assumptions on the density increase rate, but the calculated current density was insensitive to these variations. The value for the initial on-axis free electron density is taken as the average measured free electron density during the first 1.5 ms after the argon valve trigger, since this is the time period before the measured chord-length-averaged density starts to increase due to the argon injection. The resulting time evolution of the free electron density is shown in figure \ref{fig:density}a, along with the measured free electron density.

\subsubsection{Argon density}
\label{sec:argon}
The injected amount of argon is quantified by the pressure $p_{\rm{Ar}}$ in the MGI chamber holding the argon gas before injection. This quantity is listed in table \ref{tab:discharges}, as well as the corresponding number of injected argon atoms $N_{\rm{Ar}}$, assuming a valve volume of 0.1 l and a temperature of 300 K. The average argon density inside the tokamak vacuum chamber after the injection can be calculated, but does not necessarily equal the on-axis argon density. Thus, some assumptions have to be made regarding which fraction of the injected argon assimilates in the plasma (referred to as the assimilation factor, $f$), and also the time dependence of the assimilation. $f$ is defined as the fraction of the total injected argon which, after the MHD mixing, resides within the plasma region defined by the major and minor radii listed in table \ref{tab:AUG}.

\begin{figure}
(a) \subfloat{\includegraphics[width=0.46\textwidth]{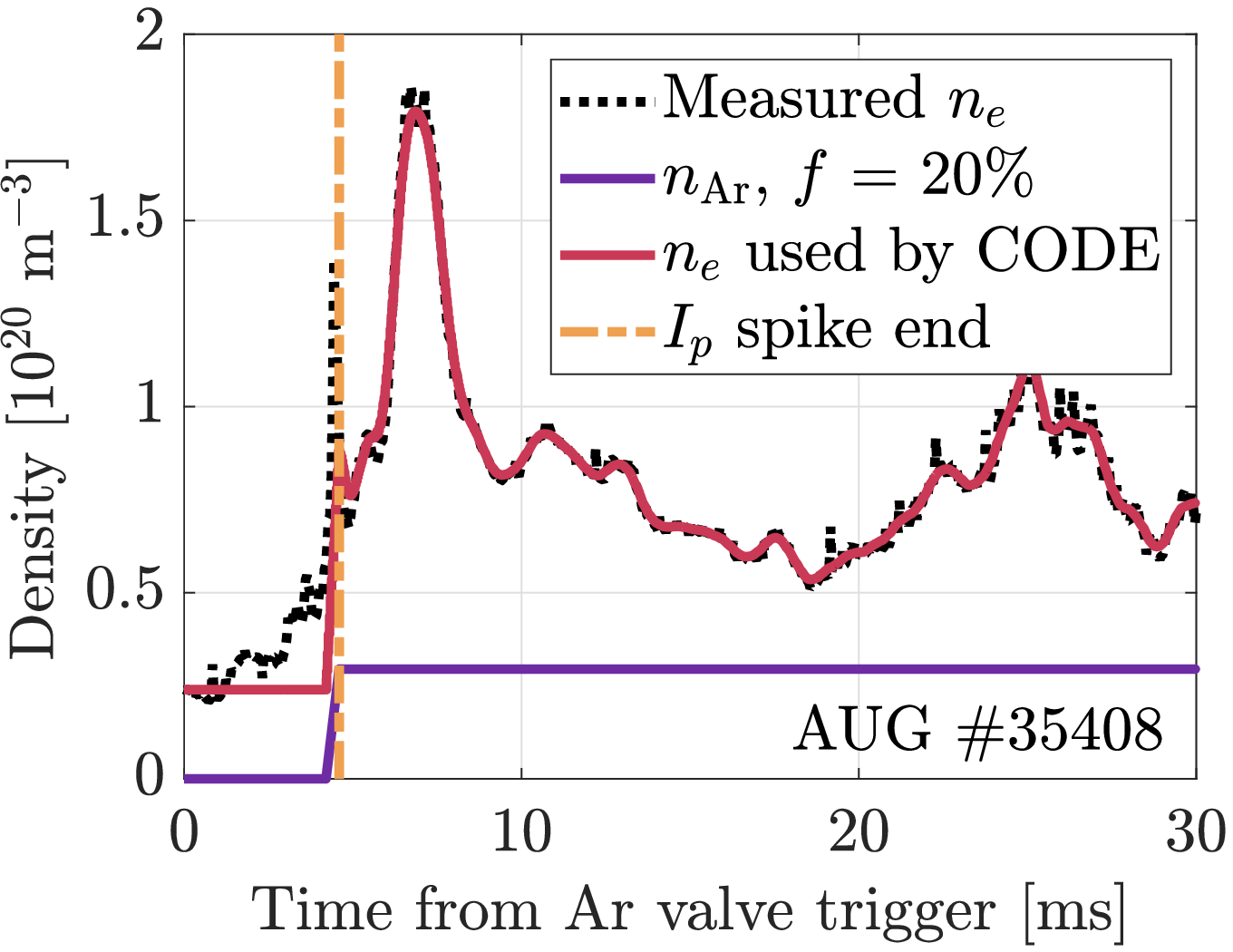}}
(b) \subfloat{\includegraphics[width=0.46\textwidth]{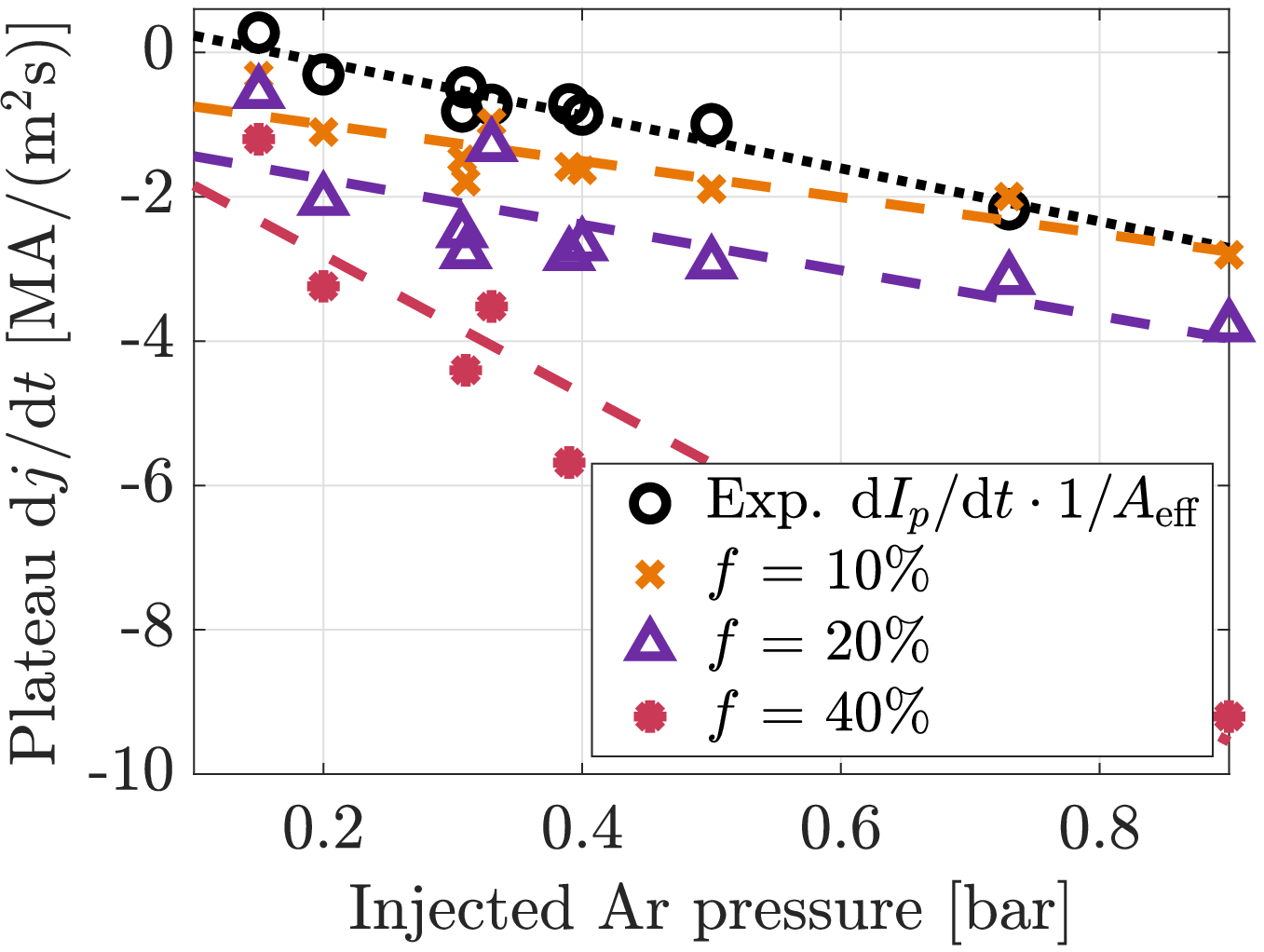}}
    \caption{(a) The free electron density measured by interferometry ($\sampleline{dotted}$) and the assumed on-axis values for the densities of argon and free electrons ($\sampleline{}$) for AUG discharge \#35408. The end of the measured $I_p$ spike is also marked with a vertical line for reference. (b) The dissipation rates for the experimentally measured total plasma current $I_p$ and the calculated current densities for three different values of the assimilation fraction $f$. Note that the $I_p$ dissipation rate has been scaled with $A_{\rm{eff}}$ = 0.63 m$^2$.}
    \label{fig:density}
\end{figure}

As will be shown later, in section \ref{sec:j}, the plasma current drops suddenly during the disruption (referred to as the \textit{current quench}, CQ) and then dissipates more slowly during the so-called plateau phase. As shown in figure \ref{fig:density}b, experimental data shows that there is a linear correlation between the injected argon amount and the current dissipation rate in this phase, as long as there is some RE generation but not full conversion of the initial current into RE current. The assimilation factor was estimated by comparison of measured and simulated current density dissipation rates, d$j$/d$t$, during the plateau phase. For comparison with the calculated current density dissipation rate, the measured current dissipation rate has been scaled with the effective area $A_{\rm{eff}}$ = 0.63 m$^2$ explained in section \ref{sec:AUG}. The current dissipation rate has been calculated, for each discharge except the no-RE discharge \#35400, from the measured current, as an average over the time period from 20 ms to 30 ms after the argon valve trigger. This time period is well into the plateau phase for all simulated discharges.

The current density dissipation rate is calculated from the current density given by the kinetic simulations as the average over the same time span as the current dissipation rate (from 20 ms to 30 ms after the argon valve trigger). The calculated dissipation rates are plotted in figure \ref{fig:density}b for $f$ = 10\%, $f$ = 20\% and $f$ = 40\%, along with a linear fit for each value of $f$. As the figure shows, the slope of the linear fit for $f$ = 20\% approximately reproduces the experimentally deduced slope. Thus, the assimilation factor $f$ = 20\% was used for all simulated discharges, as a best estimate. The argon densities given by this estimate are in agreement with results available in the literature~\citep{pautasso20generation,papp19effect}.
The positive offset of the experimental current dissipation rates comes from the fact that the runaway plateau current is in controlled ramp-down in the experiment, which in one case gives rise to a slight current increase during the early plateau phase. The electric field in the simulations for the runaway electron generation is developing self-consistently, without the comparatively small external electric field, as the inclusion of such field would require a full self-consistent simulation of the AUG automatic control system.

Similarly to the free electron density, the on-axis argon density has been shown to remain approximately constant (zero) until MHD mixing occurs at $t_{\rm onset}$ \citep{Fable-NF-16}. After $t_{\rm end}$, the argon density is assumed to be constant, at a level given by 
\begin{equation}
n_{\rm{Ar}} = \frac{N_{\rm{Ar}}f}{V_{\rm{plasma}}} = \frac{N_{\rm{Ar}}f}{2\pi^2Ra^2}.
    \label{eq:Ardens}
\end{equation}
$R$ and $a$ are the major and minor radii of the plasma given in table \ref{tab:AUG}. The assimilation fraction $f$ = 20\% was used in the simulations for all modeled discharges. Between $t_{\rm onset}$ and $t_{\rm end}$, the argon density is assumed to increase linearly. The calculated current was shown neither to be sensitive to the detailed time evolution of the density within the time interval between $t_{\rm onset}$ and $t_{\rm end}$, nor to changes of time interval length between $(t_{\rm end}-t_{\rm onset})/2$ and $(t_{\rm end}-t_{\rm onset})\cdot2$.

\subsubsection{Charge states}
\label{sec:zeff}
For the purpose of the kinetic simulations, the average charge state $Z_{\rm eff}$ of the argon is inferred from measurements, i.e. not calculated from atomic physics. $Z_{\rm eff}$ is estimated by dividing the density of free electrons attributable to argon by the density of argon atoms. All free electron density above the initial value $n_{e0}$ is attributed to argon. The corresponding distribution over the discrete ionization states is calculated by linear interpolation between the two integers closest to the calculated average charge state, \textit{e.g.} if the average charge state is calculated to be 4.5, then half of the states are assumed to be 4 and the other half 5. The thus inferred $Z_{\rm eff}$ generally shows a distinct peak of around 6 immediately after the end of the current spike and then fluctuates at lower values during the remaining simulation time.

Detailed atomic physics modelling yields a broader distribution over multiple ionization states \citep{Linder-2020}. Simulations were also performed using a distribution over multiple charge states given by equilibrium between excitation and recombination rates at the given temperature. It is however unlikely that this equilibrium is reached during the rapid thermal quench, and the difference in final calculated current was less than 1\% between the cases run with equilibrium assumption and linear interpolation respectively.

\subsection{Thermal quench} \label{sec:TQ}

The presence of impurities in the plasma abruptly increases radiative energy losses, which leads to the rapid thermal quench (TQ). The simulation results are very sensitive to the duration of the TQ.

\subsubsection{Measured free electron temperature}
\label{sec:measuredT}
Previous work by \citet{Fable-NF-16} indicates that the on-axis free electron temperature $T_e$ remains almost constant until MHD mixing occurs in MGI induced disruptions in ASDEX Upgrade. The free electron temperature is indirectly measured through electron cyclotron emission (ECE), which also yields a temperature profile in the plasma. The thus measured free electron temperature, averaged over a circular 15 cm$^2$ area surrounding the plasma axis, is shown in figure \ref{fig:temperature} ($\sampleline{dotted}$). However, the ECE signal is cut off from approximately 1.5 ms after the beginning of the argon injection until $t_{\rm end}$, due to the high free electron density at the plasma edge \citep{Fable-NF-16}. Thus, we calculate the initial free electron temperature $T_{e0}$ as the measured value averaged over the central 15 cm$^2$ area and the time interval between 1 ms and 1.5 ms after the argon injection valve trigger. This time interval was chosen to exclude both any initial temperature variations due to the heating system being shut off shortly before the disruption, and the beginning of the TQ. $T_e$ is then assumed to decrease exponentially with time $t$, as argued by \citet{Smith-PoP-08}. We thus assume that $T_e = T_{\rm exp}$, where $T_{\rm exp}$ is given by equation \ref{eq:T}, from the beginning of the simulation until $T_{\rm exp} < T_{\rm EQ}$, where $T_{\rm EQ}$ is given by equation \ref{eq:rad_equilibrium}.
\begin{equation}
	T_{\rm exp} = T_{e0} - \frac{T_{e0}-T_{e,\rm{final}}}{1+\exp\left(\frac{t_{\rm end}-t_{TQ}-t}{t_{TQ}}\right)}
    	\label{eq:T}
\end{equation}
$t_{TQ}$ is determined as described in section \ref{sec:tTQ}. The final temperature $T_{e,\rm{final}}$ can be any low value, because after the TQ, the values given by equation \ref{eq:T} are no longer used, but instead the equilibrium temperature is determined using the equilibrium assumption described in section \ref{sec:eq}. The final equilibrium temperature is approximately 1 eV for all the simulated discharges. The condition for using the equilibrium temperature is that it is higher than the temperature given by equation \ref{eq:T}, so $T_{e,\rm{final}}$ is fixed to 0.5 eV to make sure that the temperature given by equation \ref{eq:T} falls below the equilibrium temperature in all simulations.

\subsubsection{Thermal quench time}
\label{sec:tTQ}
\begin{figure}
(a) \subfloat{\includegraphics[width=0.45\textwidth]{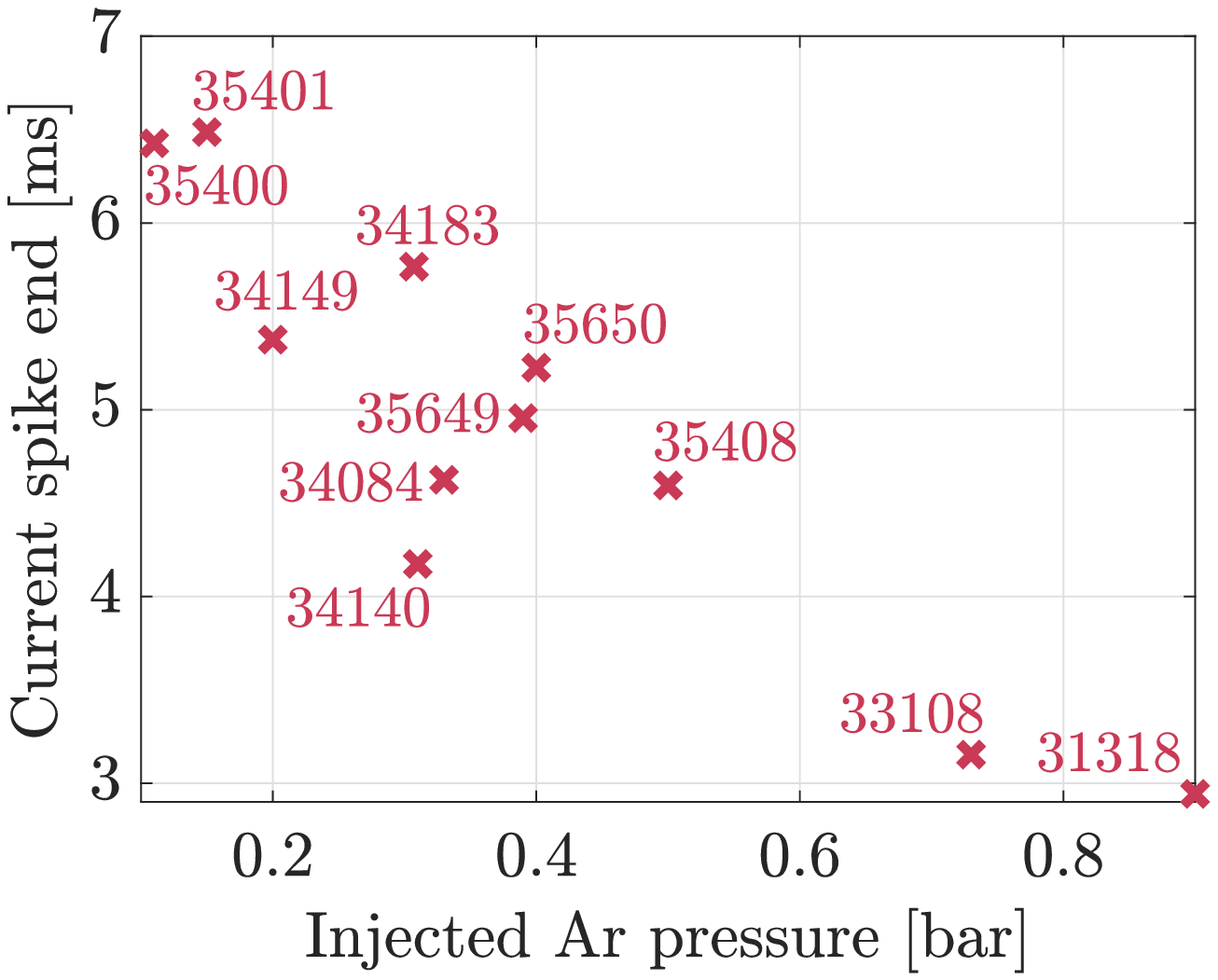}}
(b) \subfloat{\includegraphics[width=0.45\textwidth]{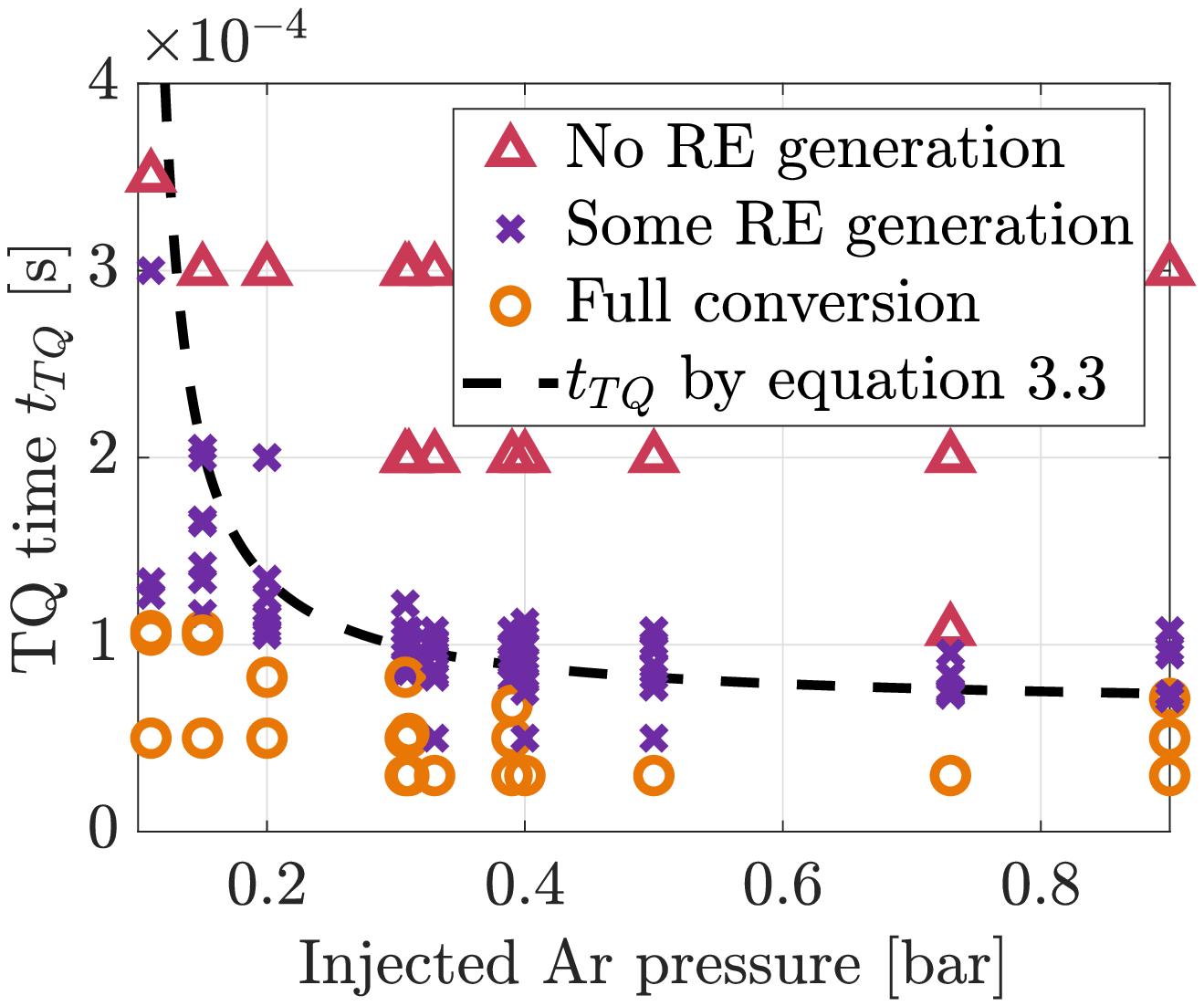}}
    \caption{(a) The timing of the current spike end, relative to the argon valve trigger time, is used as an indication of the time scale of the disruption dynamics, here plotted against $p_{\rm{Ar}}$, showing an approximately inverse relationship. (b) Assuming that the duration of the TQ also follows an inverse relationship, the corresponding parameter $t_{TQ}$ can be chosen so that none of the simulations of the discharges results in neither full conversion ($\bigcirc$) nor no RE generation ($\bigtriangleup$).}
    \label{fig:tTQ}
\end{figure}

The thermal quench time is quantified by the parameter $t_{TQ}$, whose significance is shown in equation \ref{eq:T}. At the fastest phase of the TQ, the temperature drops from 62\% to 38\% of its initial value during the time span $t_{TQ}$, as a consequence of the formulation of equation \ref{eq:T}. The choice of $t_{TQ}$ proves to be very important for the RE generation. For too large $t_{TQ}$, there is no RE generation in the simulation, whereas for too small $t_{TQ}$, the entire current density is converted to RE current density due to the exponential sensitivity of the hot-tail generation to this parameter \citep{Smith-PoP-08}. In the experiment, a RE current was formed in all modelled discharges except \#35400, but full conversion was not observed in any of the discharges, and thus $t_{TQ}$ was chosen to reproduce this result. Using the same value of $t_{TQ}$ in each discharge did not yield the desired result - a low $t_{TQ}$ resulted in no RE formation in the high-injection cases, whereas a high $t_{TQ}$ resulted in full conversion in the low-injection cases. Also, a constant value of $t_{TQ}$ would not be expected, since the time scale of the disruption is affected by, among other parameters, the number of injected argon atoms. To quantify this dependence, the time between the argon valve trigger and the end of the $I_p$ spike was studied as a function of the injected argon pressure. As shown in figure \ref{fig:tTQ}a, an inverse relationship is found between these parameters indicating that the onset of the CQ is faster for higher injected argon pressures $p_{\rm{Ar}}$. Therefore in the modelling we use the assumption that the thermal quench time is shorter for higher $p_{\rm{Ar}}$. An interpolation (fitting) formula describing the assumed relationship between $t_{\rm{TQ}}$ and $p_{\rm{Ar}}$ was used:
\begin{equation}
	t_{\rm{TQ}} = \frac{A}{p_{\rm{Ar}}-B} + C,
	\label{eq:tTQ}
\end{equation}
with $A=7 \cdot 10^{-4}$ bar$\cdot$s, $B=0.1$ bar and $C=0.65 \cdot 10^{-4}$ s. The parameters $A$, $B$ and $C$ were varied and the simulation results for the modelled discharges (represented by $p_{\rm{Ar}}$ in the respective cases) are shown in figure \ref{fig:tTQ}b, where simulations resulting in full conversion are represented by $\bigcirc$ and simulations resulting in no RE generation are represented by $\bigtriangleup$. As shown in the figure, choosing $t_{TQ}$ according to equation \ref{eq:tTQ} resulted in some RE generation in all cases except the no-RE case \#35400. In this case, $t_{TQ}$ according to equation \ref{eq:tTQ} may be unphysically long, but it has the desired effect of preventing any RE formation in the simulation. In general, our simulations indicate that $t_{TQ}$ < 0.03 ms always results in full conversion, and $t_{TQ}$ > 0.35 ms does not result in any RE generation in any of the simulated discharges. For all the presented simulations, the respective values of $t_{\rm TQ}$ were estimated using equation \ref{eq:tTQ}, inserting $p_{\rm{Ar}}$ for the respective discharge. 

\subsubsection{Post-TQ equilibrium temperature}
\label{sec:eq}

After the thermal quench and the related MHD mixing the ECE signal is still, in many discharges, blocked due to a high free electron density, and in addition the signal noise at the low post-TQ temperatures (some tens of eV) exceeds the signal by an order of magnitude in the few discharges where the density is low enough not to cut off the ECE signal. Lacking a reliable measurement, we thus need to estimate the post-TQ temperature. A reasonable estimate can be made by assuming equilibrium between ohmic heating and line radiation losses \citep{martinsolis17formation}, so that the equilibrium temperature $T_{\rm EQ}$ must satisfy:
\begin{equation}
    E^2\sigma(T_{\rm EQ},Z_\mathrm{eff}(T_{\rm EQ}))=\sum_in_\mathrm{e}(T_{\rm EQ})n_iL_i(T,n_e(T_{\rm EQ})).
    \label{eq:rad_equilibrium}
\end{equation} 
The electric field is denoted $E$, $\sigma$ is the plasma conductivity and $Z_\mathrm{eff}$ is the effective argon ion charge. The sum goes over all possible charge states $i$ of argon, $n_i$ is the density of charge state $i$ and $L_i$ is the corresponding line radiation coefficient, obtained by interpolating data from the open-ADAS database \citep{ADAS}. As before, $n_e$ and $T_e$ are the free electron density and temperature, respectively. The equation is solved iteratively. The equilibrium temperature during the plateau phase was slightly above 1 eV for all the modelled discharges.

\begin{figure}
	(a) \subfloat{\includegraphics[width=0.45\textwidth]{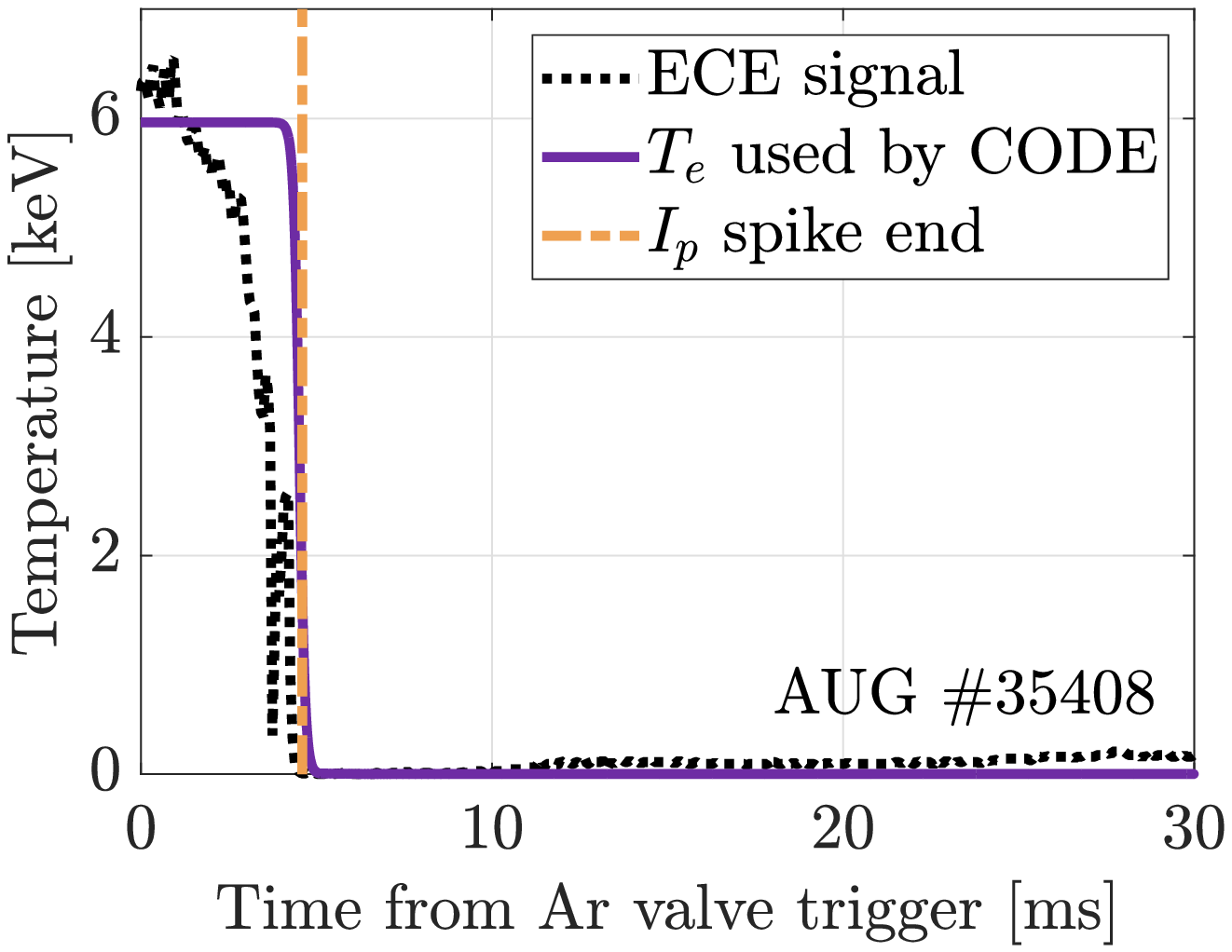}}
	(b) \subfloat{\includegraphics[width=0.45\textwidth]{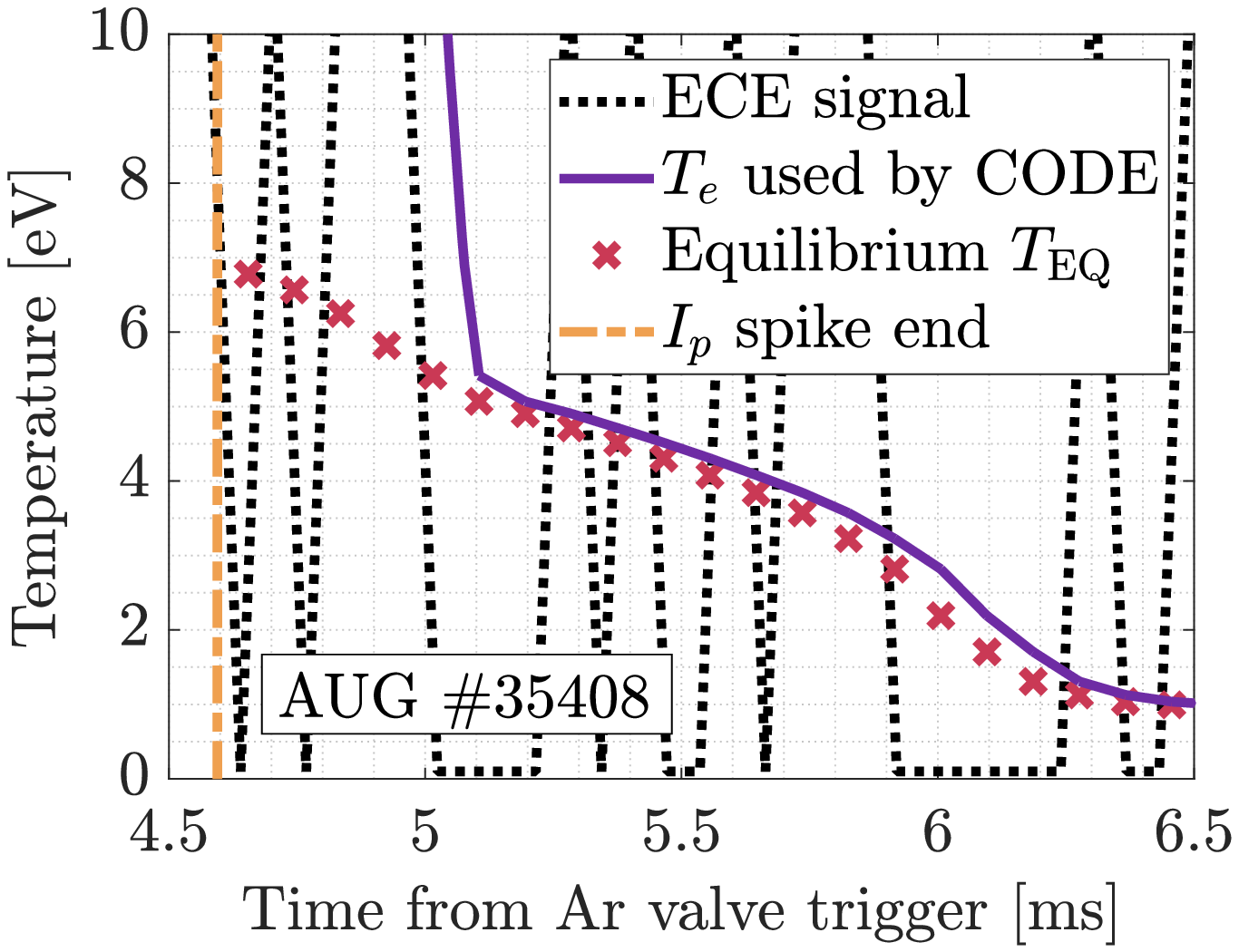}}
    	\caption{The ECE free electron temperature $T_e$ signal used in the simulations ($\sampleline{}$) and the measured temperature ($\sampleline{dotted}$) during (a) the entire simulated time span and (b) around the end of the TQ. The end of the measured $I_p$ spike is also marked by a vertical line for reference. In (b), the equilibrium temperature according to equation \ref{eq:rad_equilibrium} ($\times$) is also shown. The plot of the ECE signal in (b) demonstrates the issue with the signal noise. All data for AUG discharge \#35408.}
    	\label{fig:temperature}
\end{figure}

The calculated equilibrium temperature for discharge \#35408 is indicated with crosses in figure \ref{fig:temperature}b, which also shows the ECE-measured temperature ($\sampleline{dotted}$). As shown, the temperature used as input for the {\sc code} simulations ($\sampleline{solid}$) is taken as that given by equation \ref{eq:T} until this falls below the calculated equilibrium temperature, after which the calculated equilibrium temperature from equation \ref{eq:rad_equilibrium} is used.

\section{Results} \label{sec:results}

Using the kinetic model described in section \ref{sec:models} and experimental data as described in section \ref{sec:experimental}, the electron distribution function and associated current density is calculated. After the disruption, the ohmic current density drops rapidly in agreement with the measured $I_p$ while an RE current is formed that completely comes to dominate the total current density. Then follows a plateau phase during which the current density dissipates at a lower rate due to the interaction between the runaway electrons and the bulk plasma. For all discharges, we model the first 30 ms after the injection valve trigger, which ensures that the modeled time interval covers the entire CQ and part of the plateau phase for all the modeled cases.

\subsection{Calculated current densities}
\label{sec:j}
\begin{figure}
	(a) \subfloat{\includegraphics[width=0.45\textwidth]{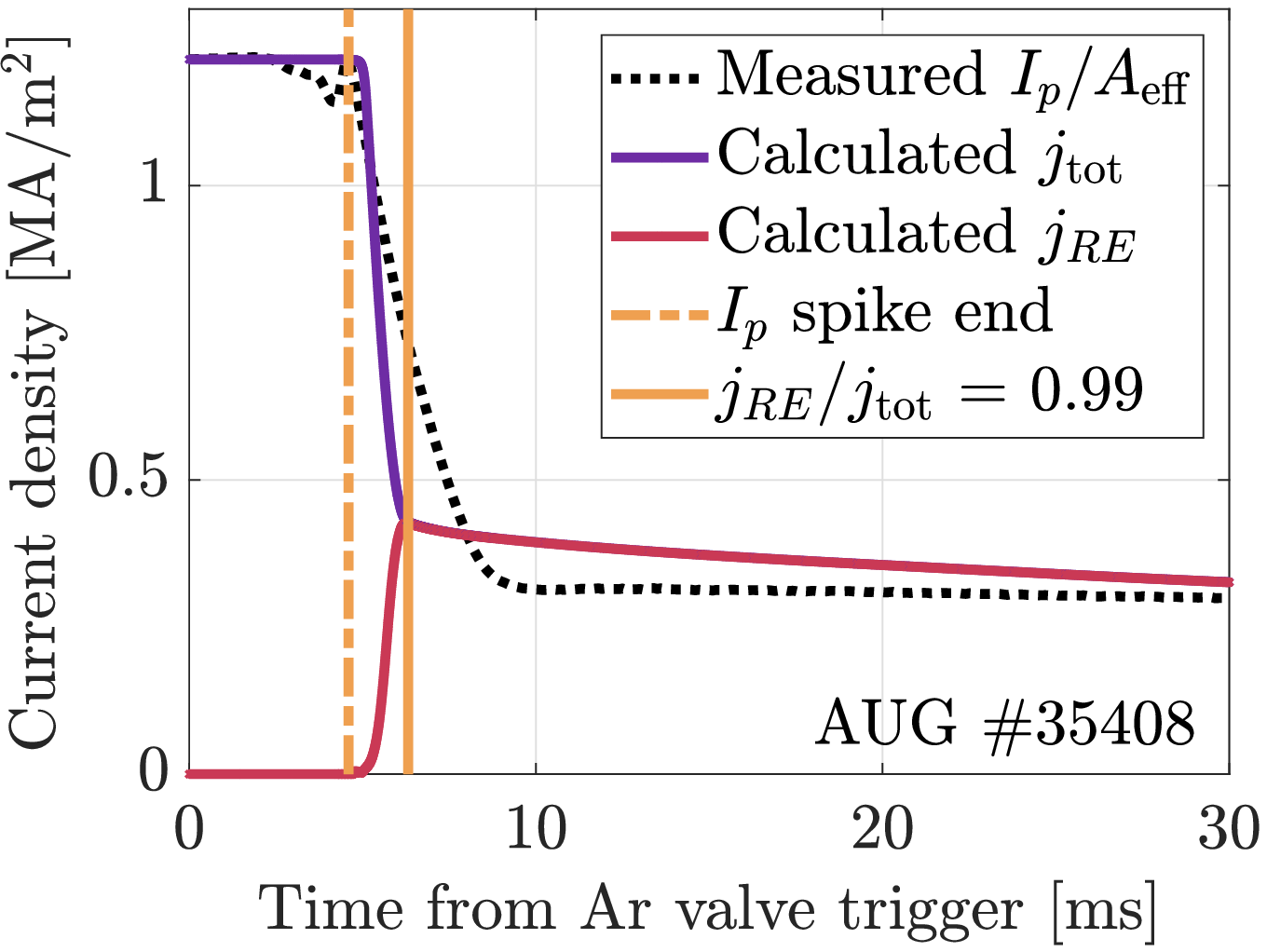}}
	(b) \subfloat{\includegraphics[width=0.45\textwidth]{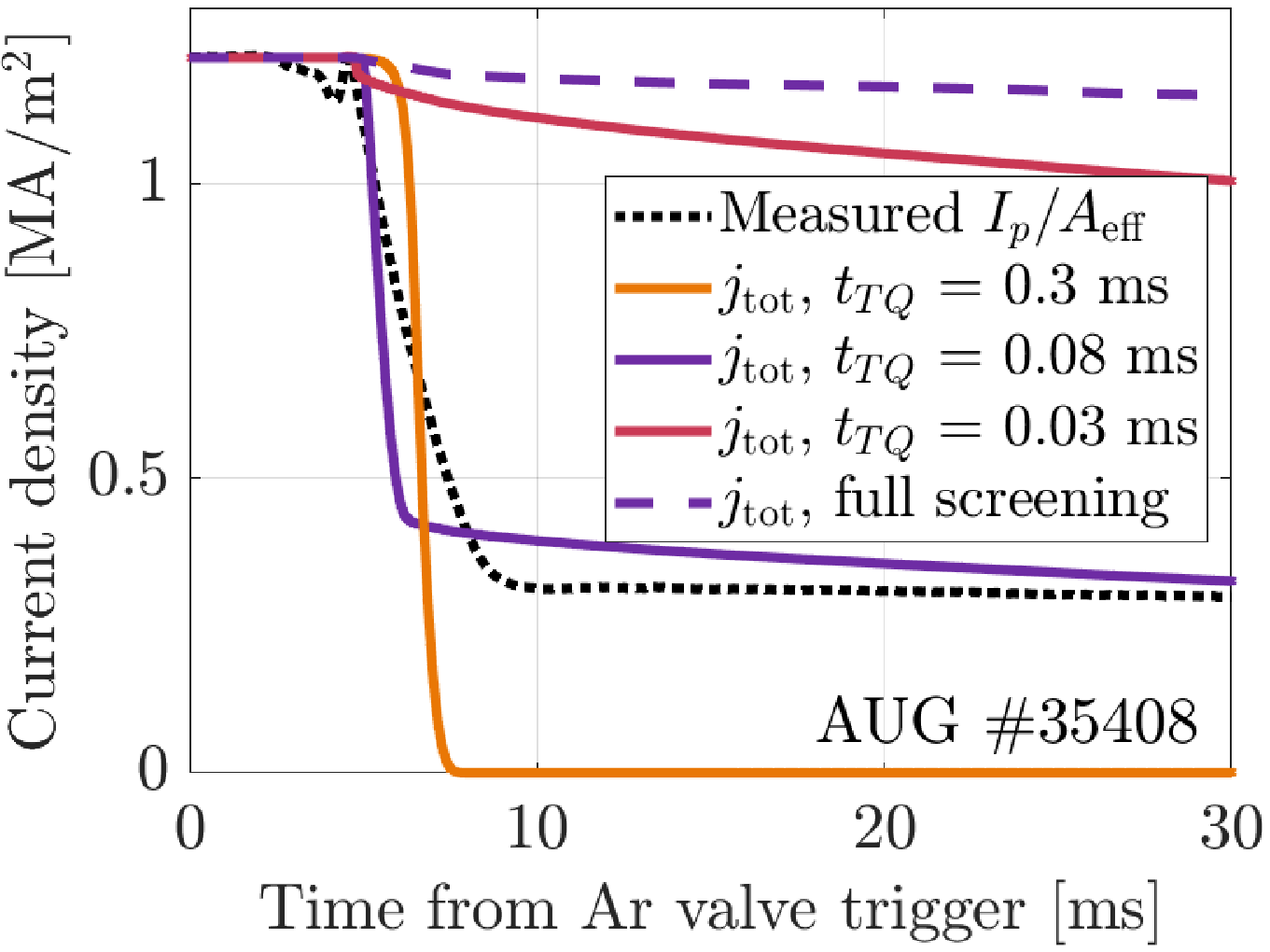}}
   	 \caption{(a) The calculated total and runaway current densities ($\sampleline{}$) are shown along with the total current divided by $A_\mathrm{eff}$ = 0.63 m$^2$ ($\sampleline{dotted}$) for AUG discharge \#35408. The end of the CQ (defined as $j_{RE}/j_{\rm tot}$ = 0.99) as marked with vertical lines. (b) The calculated total current density for \#35408 for cases with ($\sampleline{}$) and without ($\sampleline{dashed}$) screening, as well as for different values of $t_{TQ}$.}
    	\label{fig:current}
\end{figure}

The total and RE current densities are calculated by \textsc{code} by integration of the distribution function in the direction parallel to the magnetic field. These current densities are shown in figure \ref{fig:current}a, and the measured total plasma current $I_p$ is plotted for comparison, divided by the conversion factor $A_\mathrm{eff}$ = 0.63 m$^2$ that was introduced in section \ref{sec:AUG}. The motivation for this value of $A_\mathrm{eff}$, however, may not be valid for the post-disruption phase, so the comparison between the scaled total current and the calculated current density is only indicative. 

As described in section \ref{sec:tTQ}, the TQ time scale strongly affects the resulting RE generation. This fact is demonstrated in figure \ref{fig:current}b, where the calculated total current density is displayed again, and compared with the same quantity for $t_{TQ}$ = 0.03 ms and $t_{TQ}$ = 0.30 ms. For the reference case \#35408, $t_{TQ}$ = 0.08 ms. The profound effect of the screening of partially ionized impurities, mentioned in section \ref{sec:models}, is also shown in figure \ref{fig:current}b, where we also, for comparison, have plotted the total current density calculated for the reference case \#35408 with $t_{TQ}$ = 0.08 ms, but turning off the partial screening effects in the simulation, i.e. the simulation models full screening of all impurities irrespective of ionization state.

\subsection{Momentum distributions}
\begin{figure}
	(a) \subfloat{\includegraphics[width=0.45\textwidth]{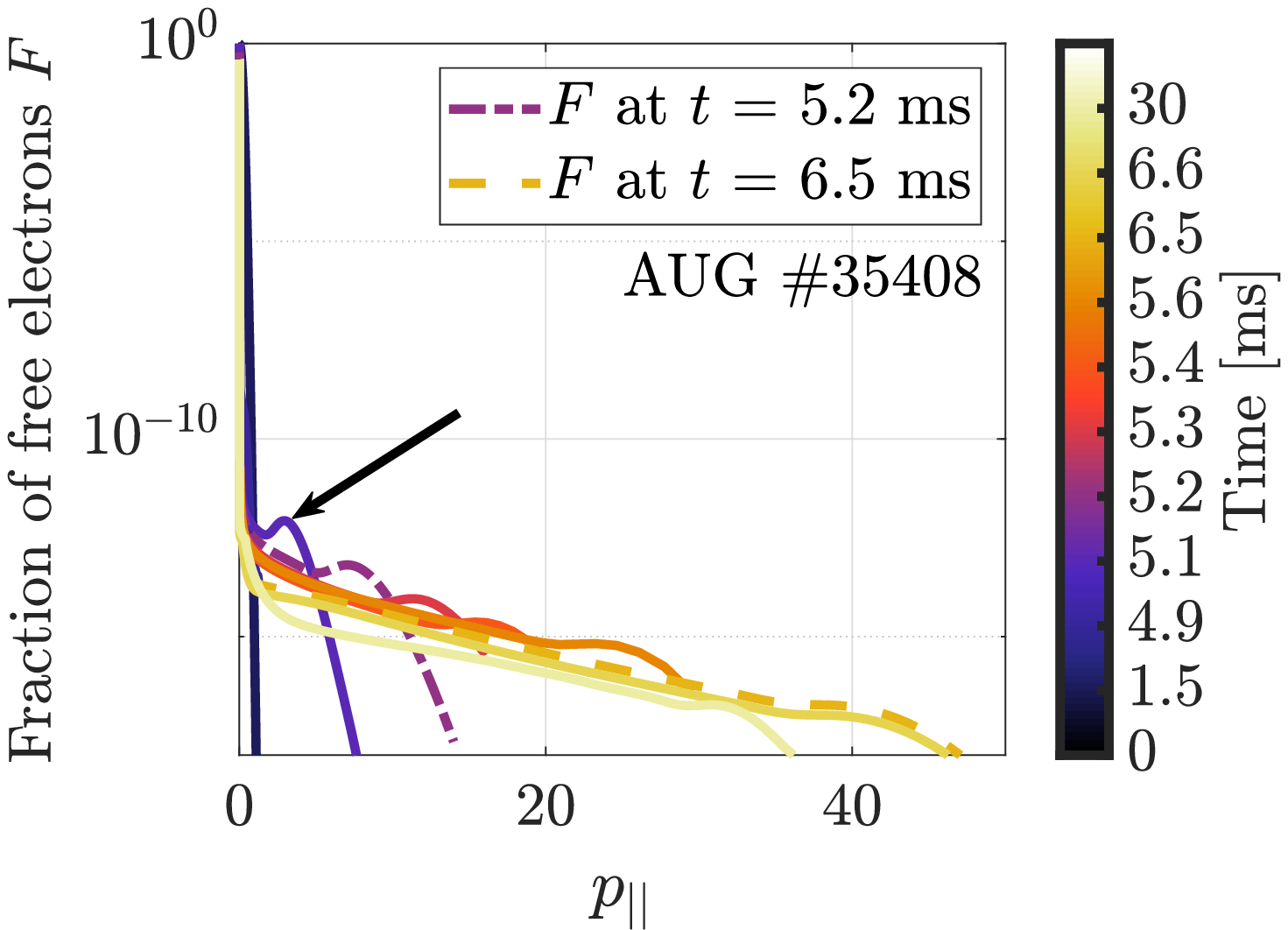}}
	(b) \subfloat{
	\begin{minipage}{0.45\textwidth}
		\includegraphics[width=\textwidth]{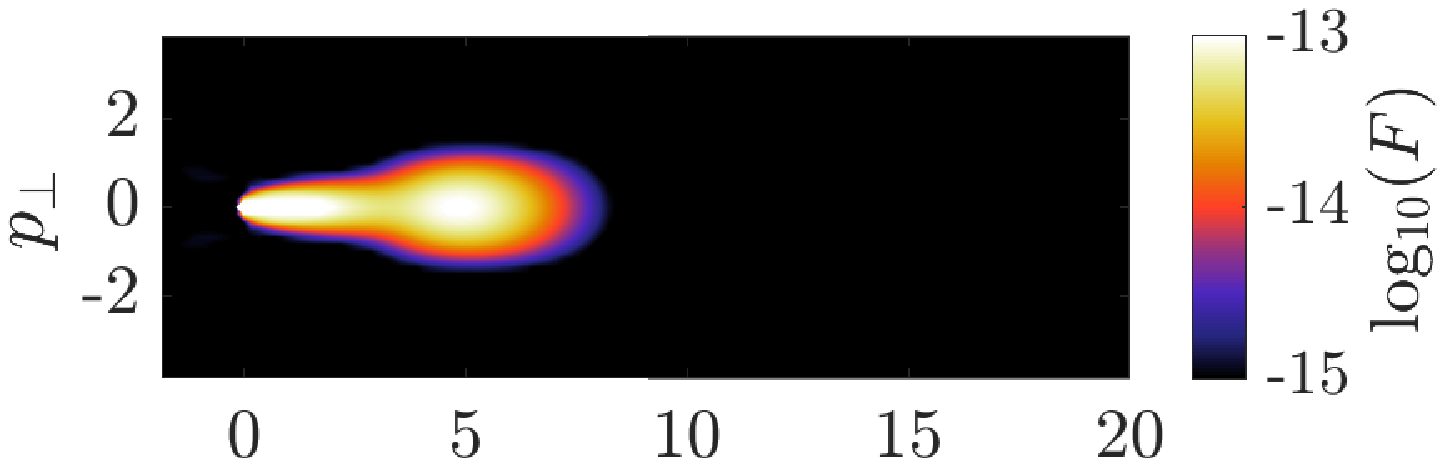}
		\includegraphics[width=\textwidth]{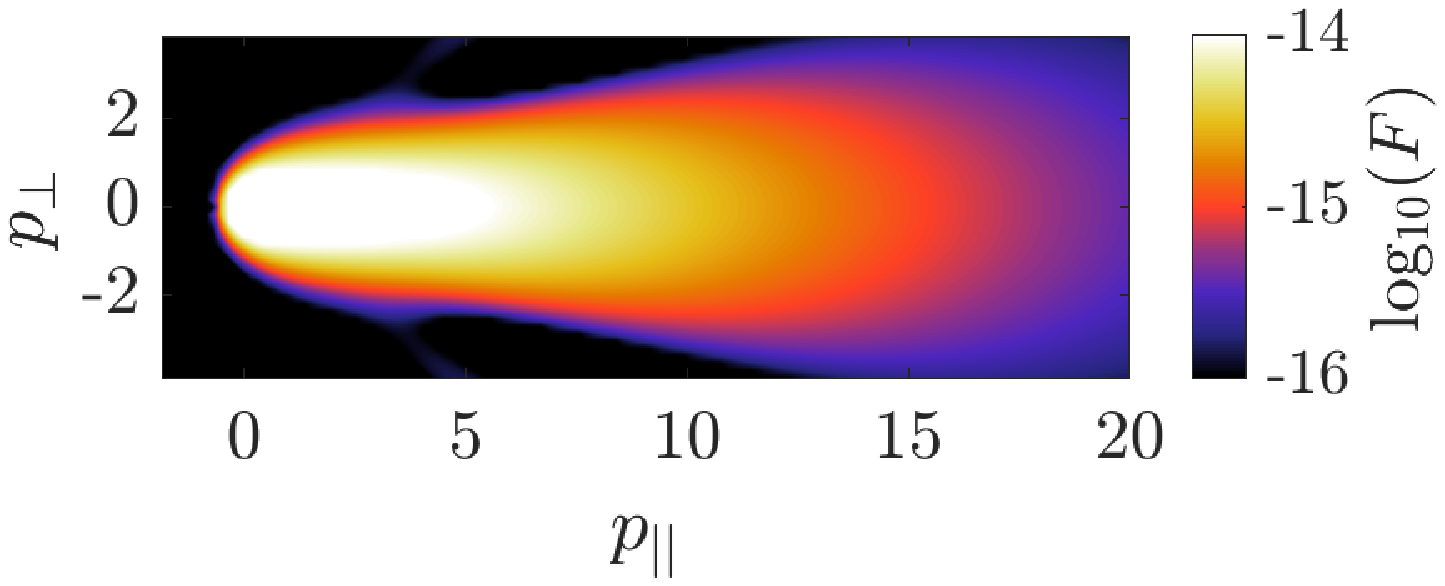}
	\end{minipage}
	}
	\caption{(a) The simulated parallel electron momentum distribution for AUG discharge \#35408. The parallel momentum distribution functions at 5.2 ms and 6.5 ms (end of CQ) are marked with \sampleline{dashdotted} and \sampleline{dashed} respectively. An arrow marks the hot-tail "bump". (b) The corresponding simulated two-dimensional electron momentum distribution at 5.2 ms (upper panel, corresponding to \sampleline{dashdotted} in figure \ref{fig:distributions}a) and at 6.5 ms (lower panel, corresponding to \sampleline{dashed} in figure \ref{fig:distributions}a). The momentum $p$ is given, in both figures, in units of $mc$, with $m$ being the electron mass and $c$ the speed of light.}
	\label{fig:distributions}
\end{figure}

The time evolution of the calculated current densities can be more thoroughly understood by observing the time evolution of the  electron momentum distribution function in the parallel (toroidal) direction. This is shown in figure \ref{fig:distributions}a for AUG discharge \#35408, where each line represents a separate time step. The non-linear scale of the color axis reflects the fact that the momentum distribution changes rapidly during the CQ but very little during the plateau phase.

During the first 5 ms, the low-momentum part of the initially Maxwellian distribution narrows due to the rapid cooling while the high-momentum tail remains almost constant, i.e.~these electrons do not lose momentum. After about 5 ms, the high-momentum tail (or "hot-tail") starts to gain momentum and forms a "bump" (marked with an arrow in figure \ref{fig:distributions}a), separate from the thermal Maxwellian which is now indistinguishable from the vertical axis. The "bump" is then gradually accelerated and simultaneously the avalanche mechanism gives rise to an approximately exponentially decreasing distribution of fast electrons at lower momenta.

A contour plot of the complete 2D momentum distribution is shown in figure \ref{fig:distributions}b, for the two time instances marked by different line styles in figure \ref{fig:distributions}a. The upper panel displays the momentum distribution when the hot-tail generated seed is most pronounced, at 5.2 ms. The lower panel displays the momentum distribution at 6.5 ms, i.e.~the end of the CQ when $j_{RE}/j_{\rm tot} = 0.99$, marked in figure \ref{fig:current}a. 

\subsection{RE generation mechanisms}

\begin{figure}
	(a) \subfloat{\includegraphics[width=0.45\textwidth]{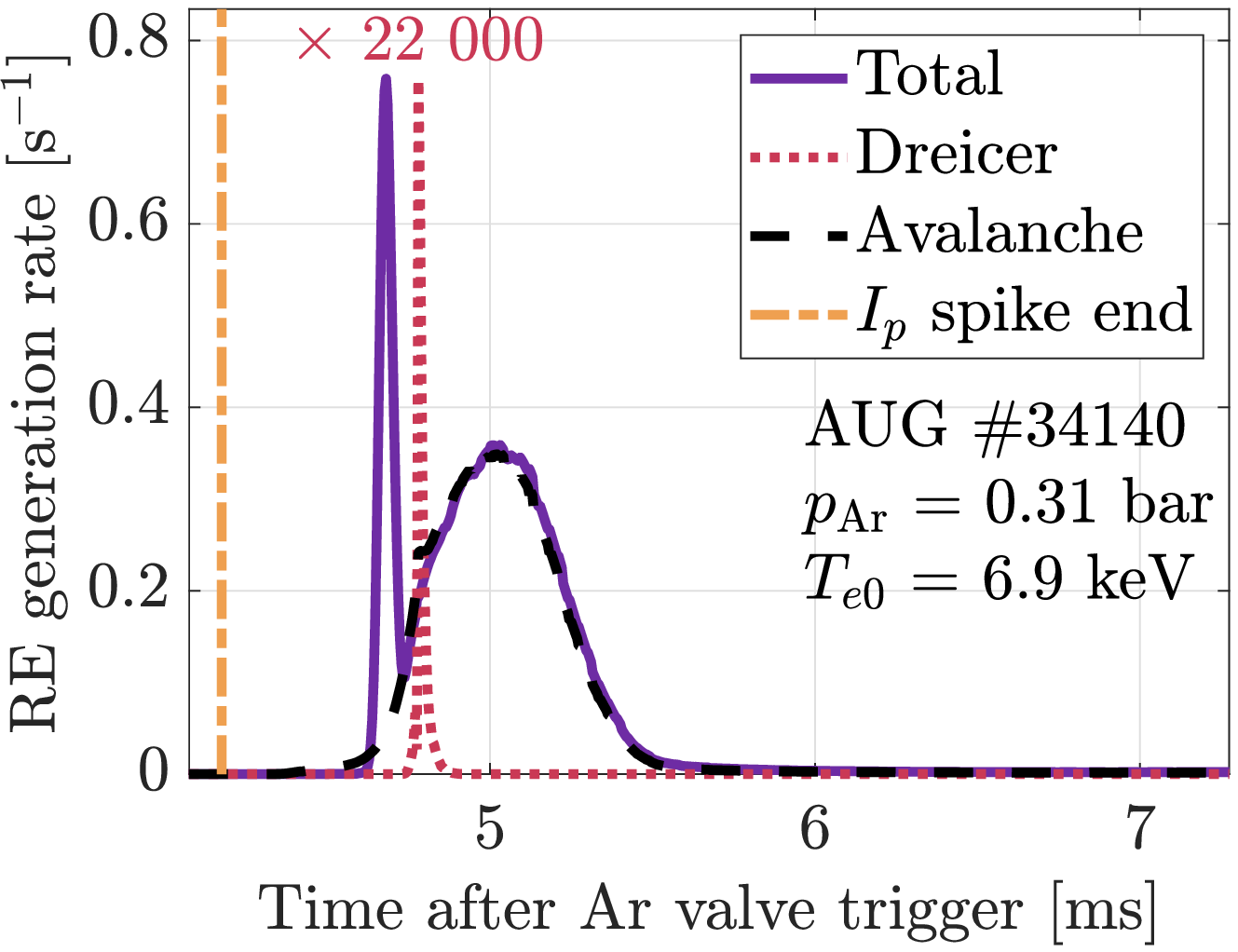}}
	(b) \subfloat{\includegraphics[width=0.45\textwidth]{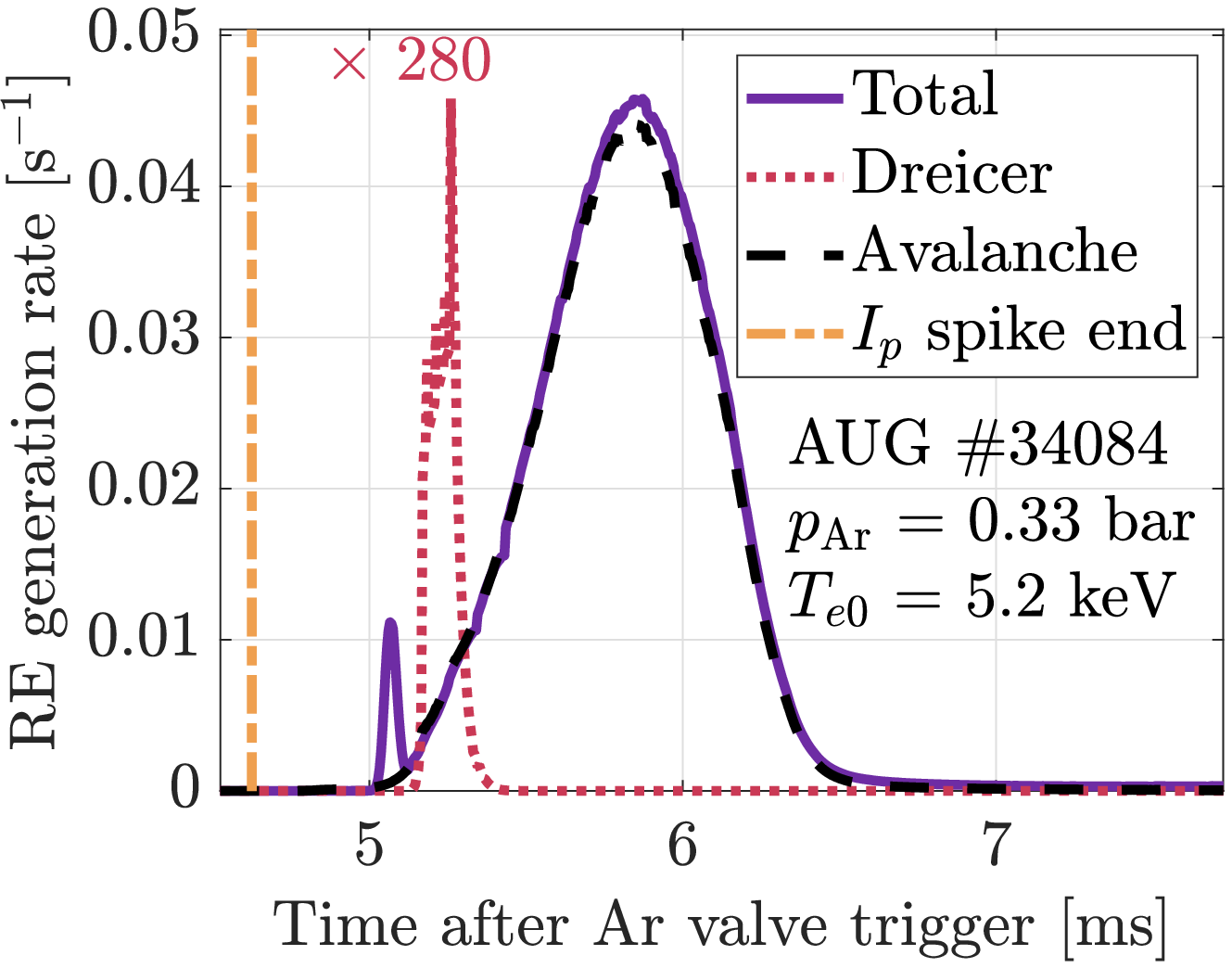}}
    	\caption{RE generation rates, total and specified per generation mechanism, for discharges \#34084 and \#34140. The Dreicer RE generation rate is scaled to be visible in the respective plots - note the different scaling factors given in the plots. The hot-tail RE seed is seen as a peak in the total RE generation rate approximately half a ms after the end of the $I_p$ spike.
}
    	\label{fig:growth}
\end{figure}

Figure \ref{fig:growth} shows the runaway generation rate $(1/n_e)({\rm d}n_{RE}/{\rm d}t)$ for two different discharges with different intial temperatures but similar injected argon pressures $p_{\rm Ar}$. Runaway electrons are, in this case, defined as having $p > 0.75~m_\mathrm{e}c$ (with $m_\mathrm{e}$ the electron rest mass and $c$ the speed of light) and the total RE generation rate is directly derived from the simulation output as the increase in the fraction of RE electrons during each time step, divided by the length of the time step.

\textsc{code} calculates the evolution of the electron distribution, and as such does not categorize REs by generation mechanisms. For a detailed analysis of the RE current formation mechanisms, the Dreicer RE generation rate was also calculated using a neural network described in \citet{Hesslow-JPP-19}, and the avalanche growth rate was also calculated using the semi-analytical formula developed in \citet{Hesslow-NF-19}. As shown in figure \ref{fig:growth}, the analytically calculated avalanche growth rate describes the total simulated growth rate well, except for in the beginning, where the hot-tail seed can be seen as a peak approximately half a ms after the end of the $I_p$ spike.

\begin{figure}
	\begin{centering}
	\includegraphics[width=0.45\textwidth]{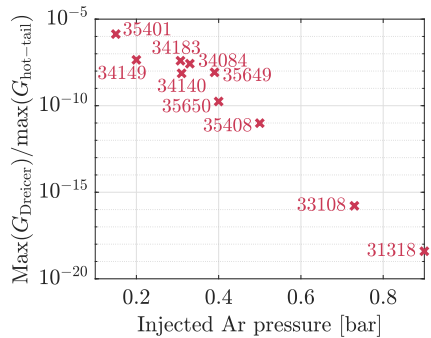}
    \caption{The ratio between the maximal Dreicer RE seed generation rate $G_{\rm Dreicer}$ and the maximal hot-tail RE seed generation rate $G_{\rm hot-tail}$, as a function of injected Ar pressure $p_{\rm Ar}$.}
    \label{fig:maxDreicerOverHT}
	\end{centering}
\end{figure}

The Dreicer RE seed generation is seen as a peak immediately after the hot-tail peak. This timing is expected since the hot-tail generation is directly connected with the TQ whereas the Dreicer generation is a consequence of the electric field generated, with some delay, after the TQ. The Dreicer generation rate was in general very small as compared with the hot-tail and avalanche generation, and smaller for larger $p_{\rm Ar}$, which is clearly illustrated by figure \ref{fig:maxDreicerOverHT}. The maximal Dreicer RE seed generation rate ($G_{\rm Dreicer}$) is never larger than 10$^{-5}$ times the maximal hot-tail RE seed generation rate, and the relative importance ranges all the way down to 10$^{-18}$ for the largest Ar injection pressure (0.9 bar, in discharge \#31318). The hot-tail RE seed generation was evaluated by subtracting the calculated avalanche generation from the total RE generation rate given by the simulations.

Generation rates for \#34084 and \#34140 are shown to display the effect of the initial temperature. These two discharges have very similar $p_{\rm Ar}$, but different initial temperatures $T_{e0}$ (5.2 and 6.9 keV respectively), and the hot-tail RE generation is found to be significantly higher for 6.9 keV, i.e. it is found to increase with increasing temperature. The Dreicer generation rates were scaled to be distinguishable in the respective plots.

\subsection{Plateau phase current and dissipation rate}
\label{sec:plateau}
\begin{figure}
(a) \subfloat{\includegraphics[width=0.45\textwidth]{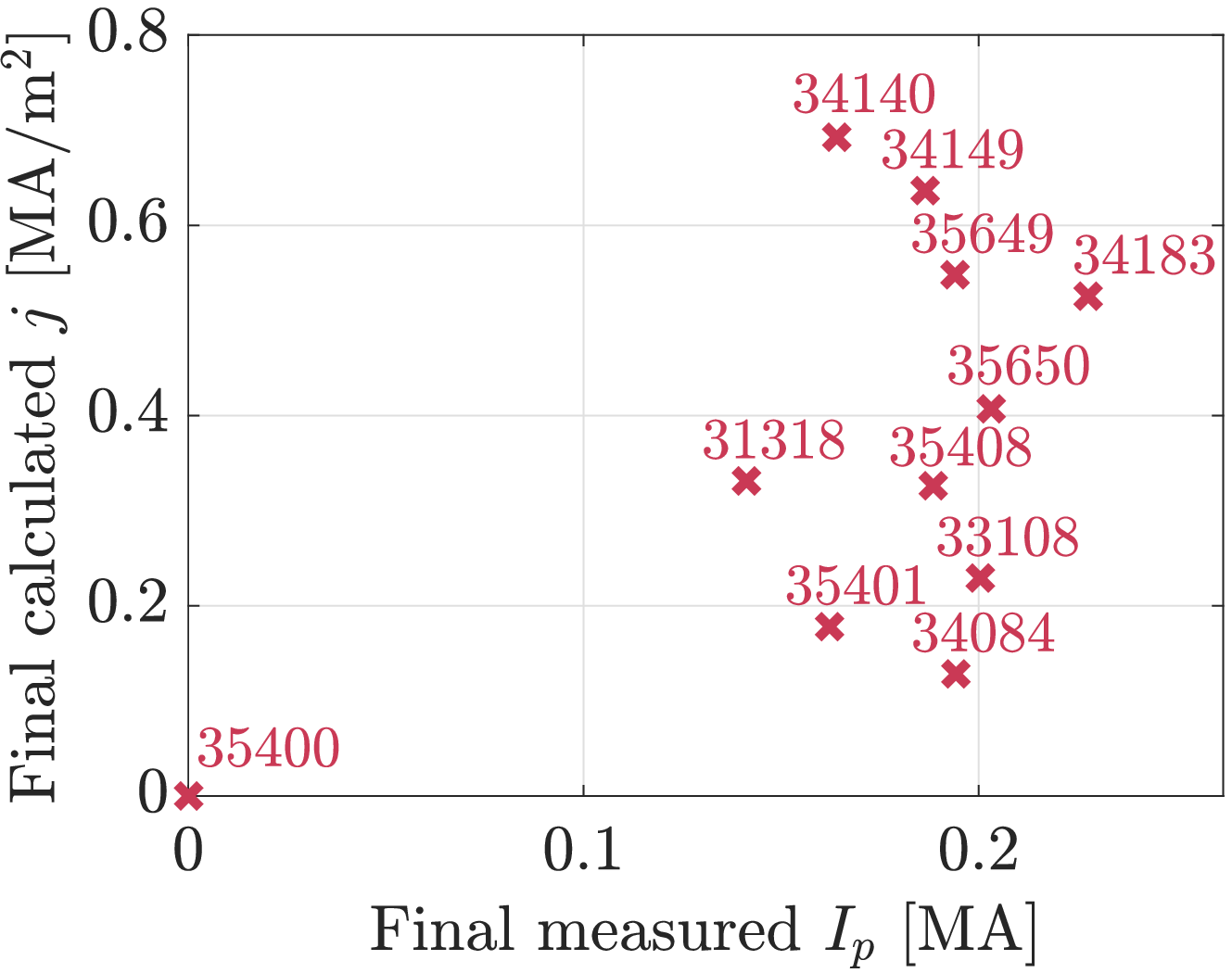}}
(b) \subfloat{\includegraphics[width=0.45\textwidth]{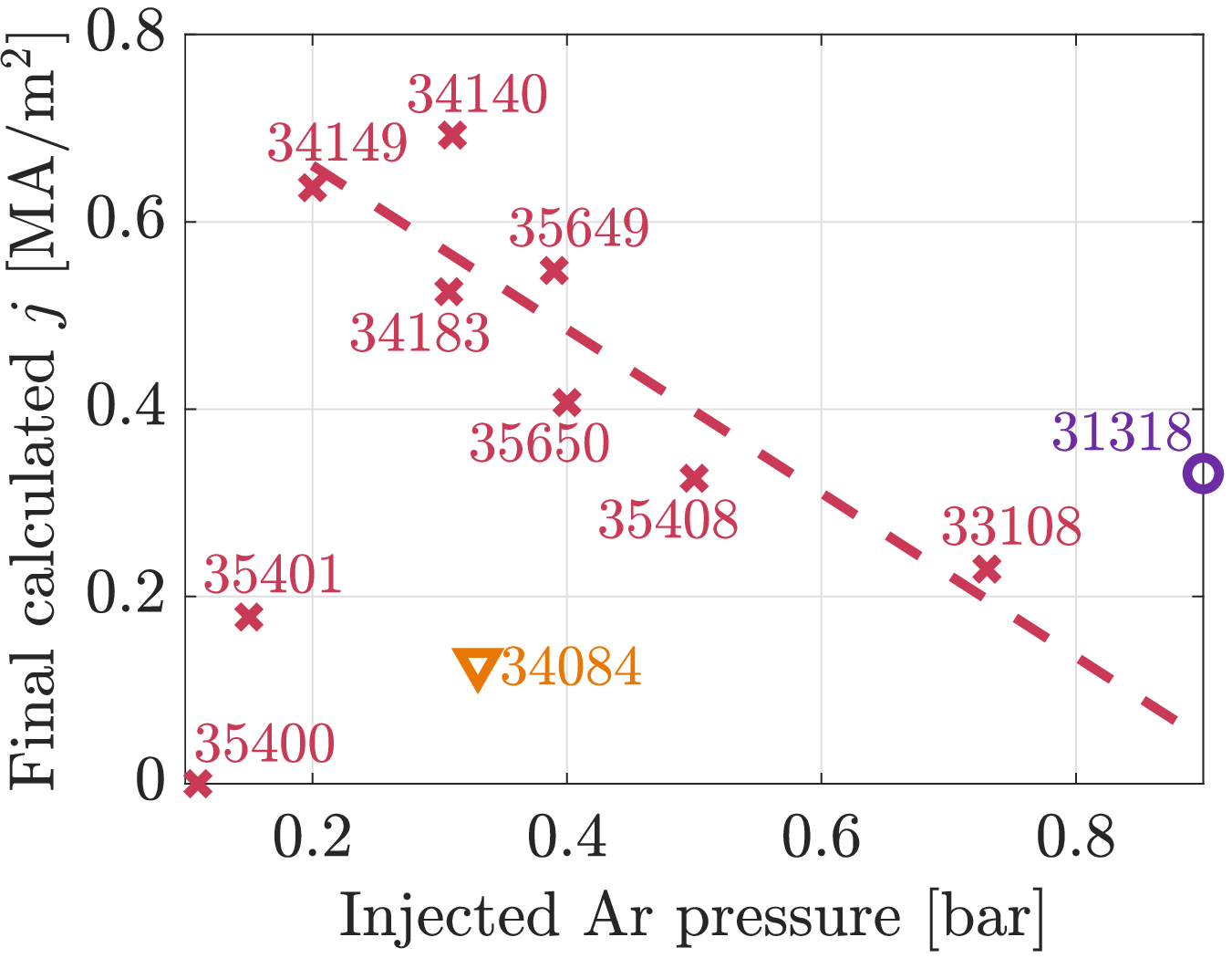}}
    \caption{The calculated post-CQ current density plotted against (a) the measured post-CQ total plasma current and (b) the injected argon pressure $p_{\rm Ar}$. In (b), a linear fit has been added to visualize the trend above the injection threshold for significant RE generation. The linear correlation excludes the outliers \#34084 ($\bigtriangledown$) and \#31318 ($\bigcirc$) as well as the discharges \#35400 and \#35401 in which $p_{\rm Ar}$ falls below the threshold.}
    \label{fig:jfin}
\end{figure}

In our simulations, the fraction of the ohmic current which is converted to RE current during the CQ is sensitive to $t_{TQ}$ and the argon assimilation factor $f$. However, on AUG the measured post-CQ plasma current shows weak correlation with individual plasma parameters such as temperature, density, or injected argon quantity; as long as these parameters are within the range of the RE generation window.
The post-CQ calculated current density is plotted against the measured post-CQ total plasma current in figure \ref{fig:jfin}a, which shows that the measured post-CQ total plasma current is fairly constant for all RE discharges, whereas the simulation results show a much larger variability and no correlation with the measurement data. This would suggest that there are negative feedback effects (such as runaway seed transport) which are not captured by the 0D kinetic modeling.

However, figure \ref{fig:jfin}b shows a correlation between the post-CQ calculated current density and $p_{\rm Ar}$. The experimental observations show that there is a certain threshold $p_{\rm Ar}$ under which no significant RE current is formed, since it does not lead to a TQ and a CQ quick enough for RE generation. Above this threshold, the post-CQ calculated current density is generally smaller for larger $p_{\rm Ar}$. Two outliers are noted, however. In discharge \#34084 ($\bigtriangledown$) the comparatively low initial free electron temperature (5.2 keV) results in a low hot-tail RE generation, as also shown in figure \ref{fig:growth}b, and hence a low post-CQ current density. In discharge \#31318 ($\bigcirc$), the initial temperature was instead comparatively high (9.3 keV), resulting in high hot-tail generation and an accordingly high post-CQ current density.

Given that we calculate a current density but only have access to an integrated total plasma current measurement, we cannot assess directly how well the simulation reproduces the on-axis current density development during the initial rapid current decrease.  The current density profile changes drastically already during the MHD phase, and thus, the previously discussed conversion factor $A_{\rm eff}$ is no longer valid. However, if we assume that the current density profile remains roughly constant throughout the plateau phase, i.e. that the conversion factor between current and current density remains constant during the plateau phase, we can make a meaningful comparison between the calculated current density dissipation rate and the measured current dissipation rate during the plateau phase. As described in section \ref{sec:argon}, it has been observed that the plateau phase current dissipation rate scales linearly with injected Ar quantity, under the condition that there is some RE generation but not full conversion of the initial current into RE current. As discussed in section \ref{sec:tTQ}, the thermal quench time parameter $t_{\rm TQ}$ was tuned to fulfill this condition, but the dissipation rate is not sensitive to $t_{\rm TQ}$ when the condition is fulfilled. As shown in figure \ref{fig:density}b, the dependence of the dissipation rate on the injected amount of Ar is reproduced by the simulations.

\section{Discussion and conclusions} \label{sec:conclusions}
We have presented kinetic modelling of runaway generation and dissipation in argon-induced disruptions in ASDEX Upgrade, where the initial on-axis free electron temperature, the free electron density, the amount of injected argon and the initial current density have been based on experimental data. The timing and fraction of the argon assimilation and the time scale of the thermal quench were estimated. The fraction of argon assimilated in the plasma volume was fixed by fitting the calculated current density dissipation rates with the experimentally observed current dissipation rates during the plateau phase. The argon was assumed to assimilate in the on-axis plasma during the MHD phase, which coincides with the spike in the measured plasma current preceding the CQ. The time scale of the TQ was assumed to be inversely proportional to the injected argon amount, similarly to the observed timing of the current spike relative to the argon injection valve trigger time. These parameters could be estimated using the same assumptions for a set of eleven discharges with varying initial temperatures and injected argon amounts, yielding reasonable simulation results for all cases, i.e. neither full conversion nor complete CQ except for the no-RE discharge \#35400. A 0D kinetic simulation cannot capture certain effects (such as e.g. runaway seed transport) and therefore the simulation results are not expected to give quantitative agreement with the experimental results.

Simulations show that, above a threshold injected argon quantity, a larger injection leads to a lower post-CQ current density, which is expected since the presence of argon increases energy loss from the plasma. The simulations also show that hot-tail RE generation is the most important RE generation mechanism in all the modelled discharges, having a significant impact on the post-CQ current density. For quantitatively accurate predictions of the plasma current, more elaborate models including transport phenomena could be used, such as the one by \cite{Linder-2020}. \textsc{code} could also be coupled with a transport code such as \textsc{go}, as done by \citet{Hoppe-ASDEX-2020}. Such simulations are underway, following up on the simulations presented herein.

    \section*{Acknowledgements}
The authors are grateful for fruitful discussions with M. Hoppe. This work has been carried out within the framework of the EUROfusion Consortium and has received funding from the Euratom research and training programme 2014 - 2018 and 2019 - 2020 under grant agreement No 633053 and from the European Research Council (ERC) under the European Union's Horizon 2020 research and innovation programme under grant agreement No 647121. The views and opinions expressed herein do not necessarily reflect those of the European Commission.  The work was also supported by the Swedish Research Council (Dnr.~2018-03911) and the EUROfusion - Theory and Advanced Simulation Coordination (E-TASC). 

\bibliographystyle{jpp}

\bibliography{plasmabibliography}

\end{document}